\documentclass[aps,superscriptaddress,nofootinbib,showpacs,eqsecnum,preprint,tightenlines]{revtex4}
\usepackage{hyperref}
\usepackage{epsfig,rotating}
\usepackage{amsmath,amssymb}
\usepackage{dsfont}

\newcommand{\be}{\begin{equation}}
\newcommand{\ee}{\end{equation}}
\newcommand{\bea}{\begin{eqnarray}}
\newcommand{\eea}{\end{eqnarray}}
\newcommand{\nn}{\nonumber}
\newcommand{\vx}{\vec{x}}

\newcommand{\vp}{\vec{p}}
\newcommand{\vP}{\vec{P}}
\newcommand{\vq}{\vec{q}}
\newcommand{\vQ}{\vec{Q}}
\newcommand{\vk}{\vec{k}}

\newcommand{\oO}{\overline{\Omega}}
\begin{document}

\title{Is the GSI anomaly due to neutrino oscillations? \\ - A real time perspective -}
  \author{Jun Wu}\email{juw31@pitt.edu}  \affiliation{Department of Physics and
Astronomy, University of Pittsburgh, Pittsburgh, PA 15260}
\author{Jimmy A. Hutasoit} \email{jhutasoi@andrew.cmu.edu}
\affiliation{Department of Physics, Carnegie Mellon University,
Pittsburgh, PA 15213, USA}
\author{Daniel Boyanovsky}
\email{boyan@pitt.edu} \affiliation{Department of Physics and
Astronomy, University of Pittsburgh, Pittsburgh, PA 15260}
\author{Richard Holman}
\email{rh4a@andrew.cmu.edu} \affiliation{Department of Physics, Carnegie Mellon University, Pittsburgh, PA 15213, USA}

\date{\today}

\begin{abstract}
We study a model for the ``GSI anomaly'' in which we obtain the time evolution of the
population of parent and daughter particles directly in real time, considering explicitly the quantum entanglement between the daughter particle and neutrino mass eigenstates in the two-body decay. We confirm that the decay rate of the parent particle and the growth rate of the daughter particle do \emph{not} feature a time modulation from interference of neutrino mass eigenstates. The lack of interference is  a consequence of the orthogonality of the mass eigenstates. This result also follows from the density matrix obtained by tracing out the unobserved neutrino states.  We confirm  this result by  providing a complementary explanation based on Cutkosky rules applied to the Feynman diagram that describes the self-energy of the parent particle.
\end{abstract}


\maketitle

\section{Introduction}
Recent experiments at the Experimental Storage Ring (ESR) at GSI in Darmstadt have revealed an unexpected time dependent modulation in the population of parent ions ${}^{140}{\rm Pr}^{58+}$ and ${}^{142}{\rm Pm}^{62+}$ from the Electron Capture (EC) decays ${}^{140}{\rm Pr}^{58+}\rightarrow {}^{140}{\rm Ce}^{58+}+\nu_e$ and ${}^{142}{\rm Pm}^{62+}\rightarrow {}^{142}{\rm Nd}^{62+}+\nu_e$ \cite{gsi}, a phenomenon that has been dubbed the ``GSI anomaly.'' In this experiment, changes of the ions' revolution frequencies are detected by the technique of time resolved Schottky mass spectrometry. For a small number of stored ions, every decay can be resolved. Thus, a time distribution of EC decays of the parent ions can be measured. On top of the experimental decay curve, the GSI experiment observed an unexpected time modulation with a period of about $T\simeq 7s$. Such a behavior is summarized in Fig.~3-5 of \cite{gsi}.

A theoretical explanation of  this remarkable time dependent modulation of the decay rate of the parent ion suggests that it is a consequence of the interference between the neutrino mass eigenstates in the final state of the  two-body decay \cite{gsi,kienle,ivanov,faber,lipkingsi}. The authors in Refs.~\cite{kienle,ivanov,faber} argue that the total amplitude of an EC decay is a coherent sum of contributions from difference neutrino mass eigenstates. The decay probability is obtained by squaring the total amplitude, thus the interference between neutrino mass eigenstates gives rise to the observed modulation feature as a consequence of their mixing and oscillations.

\emph{If} indeed periodic modulations of EC-decay rates are a consequence of neutrino mixing, these experiments would bring an interesting alternative to long-baseline neutrino experiments for the determination of neutrino mass differences.

However, this interpretation has been re-examined and criticized in Refs.~\cite{giuntygsi,burka,kienert,peshkin} on the basis that it is not the amplitudes that must be summed coherently but the probabilities, corresponding to an \emph{incoherent} addition of the contributions from the different mass eigenstates.
This approach does not lead to any modulation as the probabilities for the decay channels into  the different mass eigenstates do not interfere.
 %%addition #1%%
A similar conclusion is reached in ref.\cite{merle} by comparing the GSI experiment to other quantum processes, both within quantum field theory and with quantum mechanical probabilities.
%%end of addition #1

The theoretical and experimental importance of understanding whether neutrino mixing and oscillations could lead to time dependent modulations in two body decays where neutrinos are a component of the final state warrants an alternative exploration of these questions.

Rather than focusing on any one of these approaches, either summing amplitudes or probabilities, in this article,  we analyze the two-body decay process \emph{differently}, by obtaining  the time evolution of the population of the parent and daughter particles and considering explicitly the entanglement between the daughter particle and neutrino mass eigenstates.

We apply the method developed in Ref.~\cite{NuDynamics10} to an analysis of the GSI anomaly, examining whether   neutrino mixing and oscillations \emph{could} be responsible for time dependent modulations in the two-body decay rate. In this approach, we obtain the kinetic equations for the populations of the parent and daughter particles directly in \emph{real time} without the necessity to invoke a coherent sum over amplitudes or a sum over probabilities, thereby bypassing the theoretical controversy.

\emph{If} the time modulation is a consequence of neutrino mixing and oscillations, then this phenomenon is robust and does not depend on the details of the parent and daughter nuclei. Therefore a simple model of charged current interactions which incorporates neutrino mixing but is  stripped off the peripheral complications of nuclear matrix elements should describe the essential physical phenomena.

Therefore,  in Section \ref{sec:model}, we introduce a model for the GSI experiment, which captures the relevant physical ingredients while neglecting all unnecessary technical complications. In Section \ref{third:number}, we obtain the time evolved state emerging from the two-body decay of the parent particle. This   is a quantum mechanically \emph{entangled state} \cite{cohen} between the daughter particle  and the neutrino, whose time evolution determines completely the number densities of parent and    daughter particles, unambiguously yielding the time dependence of their population. The result of this study confirms that interference between neutrino mass eigenstates is \emph{not responsible} for any modulation in the parent or daughter population, therefore neutrino mixing is \emph{not} the reason behind the GSI anomaly. Section \ref{fin:concl} summarizes our conclusions and comments on more recent experimental results.

\section{A Model for the GSI anomaly }\label{sec:model}
\begin{figure}[h!]
\begin{center}
\includegraphics[height=3in,width=6in,keepaspectratio=true]{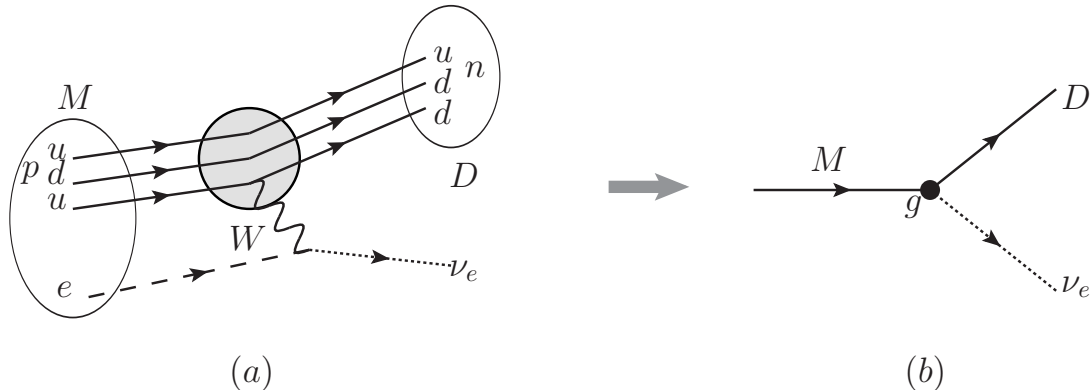}
\caption{(a) exact interaction of EC decays of a parent particle, (b) approximated interaction of EC decays in our model.} \label{fig1}
\end{center}
\end{figure}

The EC decays of heavy hydrogen-like particles are governed by charge current weak interactions, as shown in Fig.~\ref{fig1}(a).
If the observed GSI anomaly is a direct consequence of interference between different neutrino mass eigenstates in the final state as proposed in Refs.~\cite{kienle,ivanov,faber}, the technical complications associated with the details of the interaction vertices, \textit{e.g.}, spin dependence of fermionic and gauge fields, are irrelevant. In order to simplify our calculation, we introduce a bosonic model that captures the main features of these EC decays without any unnecessary complication. The two body decay now can be represented by the process shown in Fig.~\ref{fig1}(b). Our model is specified by the following Lagrangian density
\be
\mathcal{L} = \mathcal{L}_0[M, D]+ \mathcal{L}_0[\nu] +\mathcal{L}_{\rm int}[M, D, \nu_e],
\label{totallag}
\ee
with
\be {\cal L}_0[\nu] =
\frac{1}{2}\left[\partial_{\mu}\Psi^T
\partial^{\mu}\Psi  -\Psi^{T} \mathbb{M}\Psi  \right] \label{nulag}\, ,
\ee
where  $\Psi$ is a flavor doublet representing the neutrinos
\be
\Psi = \left(
             \begin{array}{c}
               \nu_e \\
               \nu_\mu \\
             \end{array}
       \right), \label{doublet}
\ee
and $\mathbb{M}$ is the mass matrix
\be \mathbb{M} = \left(
                       \begin{array}{cc}
                        m_{ee} & m_{e\mu} \\
                        m_{e\mu} & m_{\mu \mu} \\
                       \end{array}
                 \right)\,. \label{massmtx}
\ee
Here, $M$ and $D$ represent the parent and daughter particles, respectively. Their free Lagrangian density is specified by $\mathcal{L}_0[M, D]$. Also, we consider the simple case of only two neutrino flavors.

The interaction Lagrangian is analogous to the charged current interaction of the standard model, namely
\be
{\cal L}_{\rm int} (\vx,t) = g\, M(\vx,t)D( \vec{x},t)\,\nu_{e}(
\vec{x},t),
\label{Interaction}
\ee
where $g$ is the coupling constant  proportional to the Fermi constant $G_F$. We note that only electron neutrinos enter the interaction because we are considering EC decays.

The mass matrix is diagonalized by a unitary transformation
\be
U^{-1}(\theta) \, \mathbb{M}\, U(\theta) = \left(
                                                 \begin{array}{cc}
                                                  m_1 & 0 \\
                                                  0 & m_2 \\\end{array}
                                            \right)
 ~~;~~ U(\theta) = \left(
                         \begin{array}{cc}                                                                              \cos \theta & \sin \theta \\                                                                                       -\sin\theta & \cos\theta \\                                                                                     \end{array}                                                                                 \right) . \label{massU}
\ee
 In terms of the doublet of mass eigenstates,  the flavor doublet can be expressed as
\be \left(
         \begin{array}{c}
          \nu_e \\
          \nu_\mu \\
         \end{array}
    \right) = U(\theta)\,\left(
                               \begin{array}{c}
                                \nu_1 \\
                                \nu_2 \\
                               \end{array}
                          \right) \,.\label{masseigen}
\ee
In particular $\nu_e = \cos\theta~\nu_1+\sin\theta~\nu_2$.

\section{Number Densities of the Parent and the Daughter Particles} \label{third:number}

Let us consider an initial parent particle state $\big|M(\vk) \rangle$ at time $t=0$. For the GSI experiment, the parent ions are produced with a center velocity of $71\%$ of the speed of light, and with a velocity spread $\Delta v/v\simeq 5\times 10^{-7}$ \cite{gsi}.  The evolution of the number density of parent (M) and daughter (D) particles is obtained from
\bea
N_M(t) &=& \langle M(\vk)\big|\, e^{iHt}~ a^\dagger_M(\vk) a_M(\vk) ~ e^{-iHt}\,\big|M(\vk)\rangle, \nn \\
n_D(t) &=& \sum_{\vQ}n_D(\vQ,t) = \sum_{\vQ}\langle M(\vk)\big|\,e^{iHt}~ a^\dagger_{D}(\vQ) a_D(\vQ) ~ e^{-iHt}\,\big|M(\vk)\rangle, \label{numbers}
\eea
where $n_D(\vQ,t)$ is the number density of daughter particles with momentum $\vQ$. Here, the annihilation and creation operators are in the Schroedinger picture. We note that $e^{-iHt} = e^{-iH_0t}U(t,0)$ and that the number operators commute with the free field Hamiltonian. $U(t,0)$ is the time evolution operator in the interaction picture, namely,
\be
U(t,0)=T\left[e^{i\int_0^tdt'd^3x\,\mathcal{L}_{\rm int}(\vec{x},t')}\right],
\ee
where $T$ is the time-ordering operator.

Expanding $U(t,0)$ perturbatively, we obtain $U(t,0)\big|M\rangle = \big|M\rangle + \big|\Psi_D(t)\rangle^{(1)}+\big|\Psi_D(t)\rangle^{(2)}+\cdots$, where
\be
\big|\Psi_D(t)\rangle^{(1)} =  ig \int_0^t dt_1 \int d^3x_1 \Big[M(\vec{x}_1,t_1) D(\vec{x}_1,t_1)\nu_e (\vec{x}_1,t_1)\big)\Big] \big|M(\vk)\rangle, \label{psi1}
\ee
and
\bea
\big|\Psi_D(t)\rangle^{(2)} & = & - g^2 \int_0^t dt_1 \int d^3x_1 \int_0^{t_1} dt_2 \int d^3x_2 \Big[M(\vec{x}_1,t_1) D(\vec{x}_1,t_1)\nu_e (\vec{x}_1,t_1)\big)\Big]\nn \\
& & \ \ \ \ \ \ \ \ \ \ \ \ \ \ \ \ \ \ \ \ \ \ \ \ \ \ \ \ \ \ \ \ \ \ \ \ \ \ \ \ \ \ \ \Big[M(\vec{x}_2,t_2) D(\vec{x}_2,t_2)\nu_e (\vec{x}_2,t_2)\big)\Big] \big|M(\vk)\rangle, \nn \\
\label{psi2}
\eea
with $\nu_e = \cos\theta\, \nu_1 +\sin\theta\, \nu_2 $. Since $\big|\Psi_D(t)\rangle^{(1)}$ creates one daughter particle and the initial state has none, it is clear that to lowest order, the number of daughter particles is
\be
n_D(\vQ,t) = {}^{(1)}\langle \Psi_D(t)\big| a^\dagger_D(\vQ) a_D (\vQ) \big|\Psi_D(t)\rangle^{(1)} \,.  \label{enumero}
\ee
The calculation of the parent population is slightly more involved. The first order state has contributions from Fock states with zero or two  parent particles $M$, however the state with two parent particles does not conserve energy and its phase varies very rapidly in time and averages out in short time scales of order of the inverse mass of the parent particle. Therefore to obtain a non-vanishing contribution to the parent population we must consider the second order state (\ref{psi2}).

\begin{figure}[h!]
\begin{center}
\includegraphics[height=4in,width=4in,keepaspectratio=true]{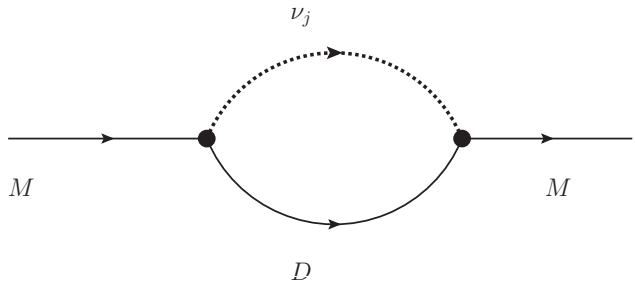}
\caption{Self-energy of the parent particle $M$, the neutrino line corresponds to a propagator of a mass eigenstate. } \label{fig:se}
\end{center}
\end{figure}

To second order, there are several contributions, but the only one that is relevant is the process
in which the first vertex at $(\vec{x_2},t_2)$ annihilates the initial $M$ creating the intermediate state with one $(D,\nu_e)$ entangled pair, while the second interaction vertex  at $(\vec{x}_1,t_1)$ \emph{annihilates} this  $(D,\nu_e)$ pair in the intermediate state and creates the $M$, which has non-vanishing overlap with $\big|M\rangle$. This process is depicted in Fig.~\ref{fig:se} and is recognized as the self-energy of the parent particle.

 Thus to lowest order in $g$, \be N_M(t) = 1+2\, \mathrm{Re}\Big[\langle M \big|\Psi_D(t)\rangle^{(2)} \Big].  \label{nWex}\ee

\subsection{Number Density of Daughter Particles: \emph{The Entangled Daughter-Neutrino State}}

As pointed out previously, for the number density of daughter particles, we only need to consider the interaction Lagrangian up to the first order, namely, Eq.~(\ref{psi1}). Expanding the field $\nu_e$ in terms of the fields that create or annihilate the mass eigenstates $\nu_1$ and $\nu_2$, and carrying out a standard quantum field theory calculation, we obtain

\bea
\big|\Psi_D(t)\big\rangle^{(1)} & \simeq & \frac{g}{\sqrt{8VE^M_{\vk} E^{D}_{\vq}}}~\sum_{\vq}\Bigg\{
     \frac{\sin\theta}{\sqrt{\Omega_{2,\vp}}}e^{i(E^M_{\vk}-E^{D}_{\vq}-\Omega_{2,\vp})\frac{t}{2}}|\nu_{2,\vp},D_{\vq} \rangle
    \Bigg[\frac{\sin\big((E^M_{\vk}-E^{D}_{\vq}-\Omega_{2,\vp})\frac{t}{2}\big)}{(E^M_{\vk}-E^D_{\vq}-\Omega_{2,\vp})/2} \Bigg]
    \nonumber \\ & & \ \ \ \ \ \ \ \ \ \ \ \ \ \ \ \ \ \ \ \ \ + \frac{\cos\theta}{\sqrt{\Omega_{1,\vp}}}e^{ i(E^M_{\vk}-E^{D}_{\vq}-\Omega_{1,\vp})\frac{t}{2}}|\nu_{1,\vp},D_{\vq} \rangle
    \Bigg[\frac{\sin\big((E^M_{\vk}-E^{D}_{\vq}-\Omega_{1,\vp})\frac{t}{2}\big)}{(E^M_{\vk}-E^D_{\vq}-\Omega_{1,\vp})/2} \Bigg]
     \Bigg\}, \nonumber \\ \label{finstate}
\eea
in which the daughter particle and the neutrinos are \emph{entangled} \cite{NuDynamics10}. Momentum conservation, a consequence of translational invariance manifest  in the Lagrangian density (\ref{totallag}), enforces $\vec{p}+\vec{q} = \vec{0}$, where $\vec{p}$ and $\vec{q}$ label the momenta of the neutrinos and the daughter particle. Also, $E^M_{\vk}$ and $E^{D}_{\vq}$ are the energies of the parent and daughter particles with momentum $\vk$ and $\vq$, respectively. $\Omega_{1,\vp}$ and $\Omega_{2,\vp}$ are the energies of the neutrino mass eigenstates with momentum $\vp$. In other words,
\be
E^M_{\vk} = \sqrt{k^2+m_M^2}, \ \ \ \ E^{D}_{\vq} = \sqrt{q^2+m_D^2}, \ \ \ \ \Omega_{1,\vp} = \sqrt{p^2+m_1^2}, \ \ \ \ \Omega_{2,\vp} = \sqrt{p^2+m_2^2}.
\ee

In order to manifestly study the time evolution of the populations and possible time dependent phenomena resulting from the mixing of mass eigenstates, we keep the finite time-dependence explicitly. As is familiar from Fermi's Golden rule, taking $t$ to infinity results in replacing
\be
\frac{\sin\big((E^M_{\vk}-E^{D}_{\vq}-\Omega_{i,\vp})\frac{t}{2}\big)}{(E^M_{\vk}-E^D_{\vq}-\Omega_{i,\vp})/2}\Bigg|_{t\rightarrow\infty} \simeq \pi\delta(E^M_{\vk}-E^{D}_{\vq}-\Omega_{i,\vp}), \label{lontime}
\ee
which leads to  the standard S-matrix result with the energy conservation at each vertex. Here $i=1,2$ stand  for  the neutrino mass eigenstates.
%If energy conservation is strictly satisfied, only one of the massive neutrinos can be produced. However, due to the finite time effect, there is always an energy uncertainty of order $1/t$. Assuming the detector is with a size of a few meters, the energy uncertainty is about $1/t\sim 10^{-7} eV$. Meanwhile, to compare $\Omega_{1,\vp}$ and $\Omega_{2,\vp}$, it is convenient to rewrite them as
%\be
%\Omega_{1,\vp} = \bar{\Omega}_{\vp}-\Delta, \ \ \ \ \ \Omega_{2,\vp} = \bar{\Omega}_{\vp}+\Delta,
%\ee
%where
%\be
%\bar{\Omega}_{\vp} = \sqrt{p^2+\frac{m_1^2+m_2^2}{2}}, \ \ \ \ \ \Delta = \frac{m_2^2-m_1^2}{4p}.
%\ee
%For the GSI experiment, $p\sim \mathcal{O}(k)\sim 100 GeV$. Therefore, the energy difference between the two massive neutrino states is give by $2\Delta \sim 10^{-15}$, much smaller than the energy uncertainty $1/t$. Consequently, such a small energy difference

It is straightforward to calculate (\ref{enumero}) with the state (\ref{finstate}). We find
\be
n_D(\vQ,t) = \frac{g^2}{8V\,E^M_{\vk} E^D_{\vQ}}\Bigg[ \frac{\cos^2\theta}{\Omega_1} \,
\frac{\sin^2\big[( E^M_{\vk} -E^D_{\vQ}-\Omega_{1})\frac{t}{2}\big]}{(E^M_{\vk} -E^D_{\vQ}-\Omega_{1})^2/4}+ \frac{\sin^2\theta}{\Omega_2} \,
\frac{\sin^2\big[( E^M_{\vk} -E^D_{\vQ}-\Omega_{2})\frac{t}{2}\big]}{(E^M_{\vk} -E^D_{\vQ}-\Omega_{2})^2/4}\Bigg], \label{ndpop}
\ee
where $\Omega_{1,2} = \sqrt{|\vk-\vQ|^2+m_{1,2}^2}$, corresponding to the specific momentum $\vQ$ of the daughter particle.

The result (\ref{ndpop}) is a consequence of the orthogonality of the Fock states associated with the mass eigenstates.

The production rate of the daughter particle is given by
\be
\frac{dn_D(\vQ,t)}{dt} = \frac{g^2}{8V\,E^M_{\vk} E^D_{\vQ}}\Bigg[ \frac{\cos^2\theta}{\Omega_1} \,
\frac{2\sin\big[( E^M_{\vk} -E^D_{\vQ}-\Omega_1)t\big]}{( E^M_{\vk} -E^D_{\vQ}-\Omega_1)}+ \frac{\sin^2\theta}{\Omega_2} \,
\frac{2\sin\big[( E^M_{\vk} -E^D_{\vQ}-\Omega_2)t\big]}{( E^M_{\vk} -E^D_{\vQ}-\Omega_2)}\Bigg].
\label{prodaughter}
\ee

The time scale of the GSI experiment is about $10-100$ seconds, corresponding to an energy uncertainty $\Delta E\sim \hbar/t\simeq 10^{-16}-10^{-17} eV$. Therefore, we can safely take the long time limit (\ref{lontime}), leading to the a \emph{constant} production \emph{rate} of daughter particles, and the total number of daughter particles produced as a function of time is given by
\be
n_D(t) = \sum_{\vQ} n_D(\vQ,t)=\Big[\Gamma_1\,\cos^2\theta + \Gamma_2\,\sin^2\theta \Big]\,t, \label{neoftinfP}
\ee
where
\bea
\Gamma_{1,2} = \frac{2\pi\,g^2}{8 E^M_{\vk}} \int
           \frac{d^3\vQ}{(2\pi)^3\,E^D_{\vQ}\,\Omega_{1,2}}
        \delta\big( E^M_{\vk} -E^D_{\vQ}-\Omega_{1,2}\big).
\eea
$\Gamma_{1,2}$ are the partial widths, while $\cos^2\theta$ and $\sin^2\theta$ are the probabilities (or branching ratios) associated with each neutrino mass eigenstate.

From the rate (\ref{prodaughter}), we see that there is \emph{no interference} between the mass eigenstates. This is a \emph{consequence of the orthogonality of the Fock states associated with mass eigenstates}. The parent particle decays through two channels, either $|\nu_1\rangle$ or $|\nu_2\rangle$ with probabilities $\cos^2\theta$ and $\sin^2\theta$ respectively,  without interference between them. Obviously when the masses of the neutrino vanish the result reduces to  the ``standard model'' decay rate, since $\Gamma_1=\Gamma_2$.

This argument becomes   clearer upon considering the process of  disentanglement of the state (\ref{psi1}). The entangled state is disentangled by the measurement  resulting in the ``collapsed" state \cite{NuDynamics10}
\bea
|\mathcal{V}_{D}(\vQ,t )\rangle & = & \frac{g}{\sqrt{8VE^M_{\vk} E^{D}_{\vQ}}}~\Bigg\{
     \frac{\sin\theta}{\sqrt{\Omega_{2}}}e^{i(E^M_{\vk}-E^{D}_{\vQ}-\Omega_{2})\frac{t}{2}}|\nu_{2,\vP} \rangle
    \Bigg[\frac{\sin\big((E^M_{\vk}-E^{D}_{\vQ}-\Omega_{2})\frac{t}{2}\big)}{(E^M_{\vk}-E^D_{\vQ}-\Omega_{2})/2} \Bigg] \nonumber \\
    & & \ \ \ \ \ \ \ \ \ \ \ \ \ \ \ \ \ + \frac{\cos\theta}{\sqrt{\Omega_{1}}}e^{ i(E^M_{\vk}-E^{D}_{\vQ}-\Omega_{1})\frac{t}{2}}|\nu_{1,\vP} \rangle
    \Bigg[\frac{\sin\big((E^M_{\vk}-E^{D}_{\vQ}-\Omega_{1})\frac{t}{2}\big)}{(E^M_{\vk}-E^D_{\vQ}-\Omega_{1})/2} \Bigg]
     \Bigg\}, \nonumber \\
\eea
where $\vP = \vk-\vQ$. It is straightforward to confirm that
\be
n_D(t) =  \sum_{\vQ} \langle \mathcal{V}_{D}(\vQ,t ) \big|\mathcal{V}_{D}(\vQ,t )\rangle \,. \label{nid}
\ee
Because $|\nu_{1,\vP}\rangle$ and $|\nu_{2,\vP}\rangle$ are orthogonal with each other, there is no interference between these two mass eigenstates. The result (\ref{neoftinfP}) is obtained in the long time limit. \\

We can further confirm our previous result of the number density of daughter particles from the density matrix. The entangled state $\big|\Psi_D(t)\rangle^{(1)}$ is produced from the decay of a parent particle, correspondingly the density matrix describing this entangled state is
\bea
\hat{\rho}(t)& =& \big|\Psi_D(t)\big\rangle^{(1)} \big.^{(1)}\big\langle\Psi_D(t)\big| \nonumber \\
    & = &   \frac{g^2}{8 V E^M_{\vk}} \sum_{\vq}  \frac{1}{E^D_{\vq}}\Bigg\{\frac{\sin^2\theta}{\Omega_{2,\vp}} \Big|D_{\vq}, \nu_{2,\vp}\rangle \langle D_{\vq},\nu_{2,\vp}
    \Big|\Bigg[\frac{\sin\Big(\big(E^M_{\vk}-E^D_{\vq}-\Omega_{2,\vp}  \big)\frac{t}{2}\Big)}{\big(E^M_{\vk}-E^D_{\vq}-\Omega_{2,\vp}\big)/2}\Bigg]^2 \nonumber \\
    & &  \ \ \ \ \ \ \ \ \ \ \ \ \ \ \ \ \ \ \ \ +\frac{\cos^2\theta}{\Omega_{1, \vp}}\Big|D_{\vq}, \nu_{1,\vp}\rangle \langle D_{\vq},
    \nu_{1,\vp}\Big|\Bigg[\frac{\sin\Big(\big(E^M_{\vk}-E^D_{\vq}-\Omega_{1,\vp}  \big)\frac{t}{2}\Big)}{\big(E^M_{\vk}-E^D_{\vq}-\Omega_{1,\vp}\big)/2 }\Bigg]^2 \nonumber \\
    & &  \ \ \ \ \ \ \ \ \ \ \ \ \ \ \ \ \ \ \ \ +\frac{\sin2\theta}{2\sqrt{\Omega_{2, \vp}~\Omega_{1,\vp}}}
    \Bigg[\frac{\sin\Big(\big(E^M_{\vk}-E^D_{\vq}-\Omega_{2,\vp}  \big)\frac{t}{2}\Big)}{\big(E^M_{\vk}-E^D_{\vq}-\Omega_{2,\vp}\big)/2 }
    \Bigg]\Bigg[\frac{\sin\Big(\big(E^M_{\vk}-E^D_{\vq}-\Omega_{1,\vp}  \big)\frac{t}{2}\Big)}{\big(E^M_{\vk}-E^D_{\vq}-\Omega_{1,\vp}\big)/2 }\Bigg]\nonumber \\
    & & \ \ \ \ \ \ \ \ \ \ \ \ \ \ \ \ \ \ \ \ \ \ \ \ \ \ \ \ \ \ \ \ \ \ \ \ \ \times\Bigg[ e^{-i\frac{\delta m^2}{4\oO} t}\Big|D_{\vq}, \nu_{2,\vp} \rangle \langle
    D_{\vq}, \nu_{1,\vp}\Big|+ e^{ i\frac{\delta m^2}{4 \oO} t} \Big|D_{\vq}, \nu_{1,\vp}\rangle \langle
    D_{\vq}, \nu_{2,\vp}\Big|\Bigg] \Bigg\}, \nonumber \\
\eea
where $\delta m^2 = m_2^2-m_1^2$, and $\bar{\Omega} = \sqrt{p^2+(m_2^2+m_1^2)/2}$ is the average energy. The density matrix contains both diagonal terms,  which describe the time evolution of the populations of the neutrino mass eigenstates, and off-diagonal terms, which display their interference\cite{NuDynamics10}.  In the GSI experiment, these neutrinos are not measured, therefore, to calculate the number density of daughter particles, we trace out the neutrino states, namely
\be
n_D(t) = {\rm Tr}_{\nu_j}\left[\hat{\rho}(t)\sum_{\vQ}a^{\dagger}_D(\vQ)a_D(\vQ)\right].
\ee
Only diagonal terms contribute to the trace because of the orthogonality of the neutrino mass eigenstates. Therefore, the number density of daughter particles has nothing to do with the interference between neutrino mass eigenstates, which is manifest in the off diagonal density matrix elements (coherence). This is consistent with the arguments in \cite{kienert} stating that the GSI experiment must be described by an incoherent sum over different neutrino states. Once again, the answer is Eq.~(\ref{prodaughter}) in the long time limit.

\subsection{Number Density of Parent Particles}
Now, let us consider the number density of the parent particles, which follows the same line of argument. It proves more convenient to calculate $dN_M(\vk,t)/dt$, for which we find\footnote{Effectively, we are obtaining the Boltzmann equation for the parent particle, neglecting the build-up of the population.}
\be
\frac{dN_M(\vk,t)}{dt}= -\frac{g^2}{4 E^M_{\vk}} \int
\frac{d^3\vQ}{(2\pi)^3\,E^D_{\vQ}} \Bigg\{ \frac{\cos^2\theta}{\Omega_{1}}~
\frac{ \sin\Big[\big( E^M_{\vk} -E^D_{\vQ}-\Omega_{1}\big)t\Big]}{\big( E^M_{\vk} -E^D_{\vQ}-\Omega_{1}\big)} +\frac{\sin^2\theta}{\Omega_{2}}~
\frac{ \sin\Big[\big( E^M_{\vk} -E^D_{\vQ}-\Omega_{2}\big)t\Big]}{\big( E^M_{\vk} -E^D_{\vQ}-\Omega_{2}\big)} \Bigg\}.  \label{dnedt}\ee
In the long time limit this becomes
\be \frac{dN_M(\vk,t)}{dt} = -\Big[\Gamma_1\,
\cos^2\theta   + \Gamma_2\,\sin^2\theta  \Big].
\label{Wrate}
\ee
Clearly, $dN_M(\vk,t)/dt = - dn_D(t)/dt$ as the decay of the parent population results in the growth of the daughter population with the same rate. This is a consequence of unitarity and we can see this by substituting (\ref{nid}) and (\ref{nWex}) into the unitarity condition
% change #2
\be
1 = \langle M(\vk)\big|U^\dagger(t,0)U(t,0)\big|M(\vk)\rangle = 1+ {}^{(1)}\langle \Psi_D(t)\big|  \Psi_D(t)\rangle^{(1)}+ 2\, \mathrm{Re}\Big[\langle M \big|\Psi_D(t)\rangle^{(2)}\rangle \Big] +\mathcal{O}(g^3)+\cdots \label{unitarity}
\ee

% end change #2

Although we have used plane waves to describe our parent and daughter particles our main result, the lack of interference of mass eigenstates in the final state is a direct consequence of the orthogonality of the mass eigenstates, and this generalizes straightforwardly to the case of wave packets. In particular, in reference~\cite{ivanov}, the wave-packet aspect of the parent and daughter nuclei is emphasized as an important ingredient to allow the neutrino mixing.

However, it is straightforward to show how the main results generalize to the case of wave-packets: a wave-packet is a superposition of   plane wave components, namely,
\begin{equation}
|\Psi(\vec{X}_0, \vec{P}_0; \vec{x}, t=0)\rangle = \int d^3p\, f(\vec{x}, \vec{X}_0; \vp, \vec{P}_0)\,e^{i\vp \cdot\vec{x}}|\vp\rangle
\end{equation}
at an initial time $t=0$. Here, $\vec{X}_0$, $\vec{P}_0$ are the center position and momentum of wave packet, respectively, while $\vec{x}$, $\vp$ and $E_p$ are the position, momentum and energy, respectively. The function $f(\vec{x}, \vec{X}_0; \vp, \vec{P}_0)$ specifies the wave function of the particle.

In our calculation, we obtain the time evolution of each plane wave component $|\vp\rangle$ from which it follows that,
\be
n_{\Psi}(t) = \int d^3p \left|f(\vec{x}, \vec{X}_0; \vp, \vec{P}_0)\right|^2~n_{\Psi}(\vp, t), \label{number:wp}
\ee
where $n_{\Psi}(\vp, t)$ is the parent or  daughter population for plane waves obtained above. The distribution function $f$ just weights each plane wave component. As demonstrated by our calculation in section \ref{third:number}, the populations $n_\Psi(\vp,t)$ \emph{do not feature oscillations} because of the orthogonality of the mass eigenstates. Therefore, interference between different neutrino mass eigenstates does not appear either in the wave packet treatment, as shown in (\ref{number:wp}). Again, this is a consequence of the orthogonality of neutrino mass eigenstates.  Obviously, this result is independent of whether the parent or daughter particles are described by plane waves or wave packets  in agreement with   reference\cite{goldman2}.

A complementary pathway to the same conclusion is provided by the interpretation of (\ref{Wrate}) in terms of the Feynmann diagram depicted in Fig.~\ref{fig:se}. This also manifestly leads to the conclusion of lack of interference between mass eigenstates because the decay rate of the parent particle is the imaginary part of the self-energy. Since the correct propagating degrees of freedom are the neutrino \emph{mass eigenstates}, the total self energy is the \emph{sum} of  self-energies with the neutrino mass eigenstates in the intermediate state. As a result, the usual Cutkosky rules indicate that the total decay width is the \emph{sum} of the partial decay widths into the mass eigenstates without interference. The real time calculations of the decay and production rates  presented above confirm this result directly from the evolution of the parent and daughter populations.
%%addition # 2 %
Introducing a complete set of intermediate states between   $U^\dagger(t,0)$ and $U(t,0$ in  the unitarity condition (\ref{unitarity}) it is straightforward to find the same result as obtained from Cutkosky rules,
directly in real time.
%%end of addition #2 %%

Thus, we confirm the analysis of Refs.~\cite{giuntygsi,kienert,burka,peshkin} that there is \emph{no interference} of mass eigenstates and we conclude that the GSI anomaly \emph{cannot} be explained in terms of the interference of mass eigenstates in the decay.

\section{Conclusions and Discussions} \label{fin:concl}

In this article, we re-examine the GSI anomaly within a framework that is  different from and complementary to previous work of various groups \cite{kienle,ivanov,faber,lipkingsi, giuntygsi,burka,kienert,peshkin}. The  controversy in the literature on the theoretical analysis of the GSI anomaly mainly focuses on whether probabilities must be summed incoherently \cite{giuntygsi,burka,kienert,peshkin} or amplitudes must be summed coherently \cite{kienle,ivanov,faber,lipkingsi}. We offer a completely different alternative to study this phenomenon: we obtain directly the time evolution of the population of parent and daughter particles taking into account that the quantum state arising from the decay of the parent particle is an \emph{entangled} state of the neutrino mass eigenstates and the daughter particle. Our method bypasses the issue of summing amplitudes or probabilities and exhibits directly the time evolution of the parent and daughter populations.

Recognizing that \emph{if} the time modulation of the parent and daughter populations is a result of interference phenomena between neutrino mass eigenstates, hence a fairly robust consequence
% change #3
is
% end change #3
independent of the complexities of the parent and daughter nuclei, we introduce a simple bosonic model that captures reliably the relevant charged current interaction process for EC and manifestly includes neutrino mixing. This allows us to extract the relevant aspects without the peripheral complications associated with spinors, nuclear wave-functions, etc. We generalize the recent work \cite{NuDynamics10} to analyze the GSI anomaly by studying the evolution of the distribution functions of the parent and daughter particles directly in time. Our starting point is the time evolution of the daughter-neutrino entangled state produced by the decay of the parent particle. This treatment also allows us to study the dynamics of the daughter particle from the density matrix upon tracing the unobserved neutrino degrees of freedom. We show that both the decay rate  of the parent particle and the production rate of the daughter particle \emph{do not feature} oscillations arising from the interference of mass eigenstates in the final state. This is a direct consequence of the orthogonality of the mass eigenstates.

  Furthermore, we provide an alternative field theoretical explanation in terms of the imaginary part of the self-energy diagram of the parent particle. The propagator of the intermediate neutrino states is that of mass eigenstates, therefore Cutkosky rules immediately lead to the conclusion that the decay rate is an incoherent sum of probabilities of decay into  each different mass eigenstate (channels), complementing and confirming our previous analysis. Simple arguments based on superposition clearly show that a wave packet treatment of parent and daughter particles yields the same result, namely no time modulation since there is no interference between mass eigenstates in the final state, again a direct consequence of orthogonality of mass eigenstates.

  Thus our work confirms the result of Refs.~\cite{giuntygsi,burka,kienert,peshkin} that if the GSI anomaly is a real effect, it cannot be explained from the interference of neutrino mass eigenstates.

  More recently, independent experimental efforts have addressed the GSI anomaly: in Ref.~\cite{gsi-2} the EC decay of $^{180}Re$ is studied and \emph{no modulation of the decay rate is observed}. However, this experiment is \emph{different} from the one at GSI not only because of the very short-lived initial state, but also more importantly because the daughter particle moves in a lattice and is restricted to transfer crystal momentum to   phonons.

Another EC-decay experiment with $^{142}Pm$   and an earlier EC-decay experiment with $^{142}Eu$  re-analyzed by Vetter \emph{et.al.} \cite{vetter} did not observe the modulation in the rates reported by the GSI experiment.

  Therefore,   our work supports the conclusion against an explanation of the GSI anomaly as a consequence of neutrino mixing in agreement with previous work \cite{giuntygsi,burka,kienert,peshkin}. These theoretical results, combined with emerging independent experimental evidence seem to suggest that if the GSI time modulation anomaly is a real phenomenon, its cause is probably associated with the details of the GSI experiment,   other mechanisms such as neutrino spin precession in the static magnetic field of the storage ring\cite{gal}, hyperfine level splitting\cite{pavli}, spin rotation\cite{spinrot} or perhaps internal nuclear degrees of freedom\cite{kennedy} of the parent particle in such an experiment, but \emph{not} a consequence of neutrino mixing.

\acknowledgements
D.~B. and J.~W. are supported by NSF grant award   PHY-0852497. J.~W. thanks support through Daniels and Mellon Fellowships. R.~H. and J.~H. are supported by the DOE through Grant No. DE-FG03-91-ER40682.
%\appendix

%\bibliographystyle{phaip}
%\pagestyle{plain}
%\bibliography{Neutrino Oscillation}

\end{document}